\documentstyle[aps,prb,multicol,graphics]{revtex}
\begin{document}
\draft
\title{Filling dependence of the Mott transition in the 
       degenerate Hubbard model}
\author{Erik Koch$^{(a,b)}$, Olle Gunnarsson$^{(a)}$,
        and Richard M.~Martin$^{(b)}$}
\address{${}^{(a)}$Max-Planck-Institut f\"ur Festk\"orperforschung,
         D-70506 Stuttgart, Germany}
\address{${}^{(b)}$Department of Physics, University of Illinois at
         Urbana-Champaign, Urbana, Il 61801}
\date{\today}
\maketitle

\begin{abstract}
Describing the doped Fullerenes using a generalized Hubbard model,
we study the Mott transition for different integer fillings of the $t_{1u}$
band. We use the opening of the energy-gap $E_g$ as a criterion for the 
transition. $E_g$ is calculated as a function of the on-site Coulomb 
interaction $U$ using fixed-node diffusion Monte Carlo. We find that for 
systems with doping away from half-filling the Mott transitions occurs at 
smaller $U$ than for the half-filled system. We give a simple model for the 
doping dependence of the Mott transition.
\end{abstract}
% 71  Electron states
% 71.10.Fd Lattice fermion models (Hubbard model, etc.)
% 71.20.Tx Fullerenes and related materials; intercalation compounds
% 71.30.+h Metal-insulator transitions and other electronic transitions

\begin{multicols}{2}

\section{Introduction}
The Hubbard Hamiltonian is a simple model for studying strongly interacting
systems. In particular it is used to investigate the Mott-Hubbard 
metal-insulator transition in half-filled systems.\cite{Mott} It is clear that 
for strong
correlations such a system should be insulating, since in the atomic limit 
the states with exactly one electron per lattice site are energetically
favored, while all other states are separated from those by a Coulomb gap.
For a generalized Hubbard model with degenerate orbitals the same argument
implies that for strong correlations not only the half-filled, but all 
integer filled systems will become Mott-Hubbard insulators. It is then natural 
to ask how the location of the transition depends on the filling.

As an example we consider a Hamiltonian describing the alkali doped 
Fullerides.\cite{c60rmp}
It comprises the three-fold degenerate $t_{1u}$ orbital and the Coulomb 
interaction $U$ between the electrons on the same molecule. Using this 
Hamiltonian, we have recently shown that, although $U$ is substantially larger
than the band width $W$, K$_3$C$_{60}$ is not a Mott insulator but a (strongly
correlated) metal.\cite{c60mott} Prompted by the synthesis of an isostructural
family of doped Fullerenes A$_n$C$_{60}$ with different fillings $n$,\cite{Fm3m}
we now address the question of the Mott transition in integer doped Fullerides.
For these systems we have the interesting situation that for fillings $n=$ 1,
2, 3, 4, and 5 calculations in the local density approximation predict them 
all to be metallic,\cite{ldabands} while in Hartree-Fock they all are 
insulators. Performing
quantum Monte Carlo calculations for the degenerate Hubbard model at different 
fillings and for values of $U$ typical for the Fullerides, we find that all 
the systems are close to a Mott transition, with the 
critical correlation strength $U_c$ at which the transition takes place 
strongly depending on the filling $n$. More generally,
our results show how, for an otherwise identical Hamiltonian, the location of 
the Mott transition $U_c$ depends on the filling. $U_c$ is largest at 
half-filling and decreases for fillings smaller or larger than half. We contrast
these findings with the results from Hartree-Fock calculations which predict
a much too small $U_c$ and show almost no doping dependence.
We give an interpretation of the results of the quantum Monte Carlo 
calculations extending the hopping argument introduced in Ref.\ 
\onlinecite{c60mott} to arbitrary integer fillings. Despite the crudeness of 
the argument it explains the doping dependence found in quantum 
Monte Carlo. We therefore believe that our simple hopping argument captures
the basic physics of the doping dependence of the Mott transition in
degenerate systems.

In section II we introduce the model Hamiltonian for doped Fullerenes with
a three-fold degenerate $t_{1u}$ band. We discuss the fixed-node approximation
used in the diffusion Monte Carlo calculations, present the results of
our quantum Monte Carlo calculations, and contrast them to the result of
Hartree-Fock calculations. Section III gives an interpretation of the 
results of our calculations in terms of intuitive hopping arguments.
We introduce the many-body enhancement of the hopping matrix elements, 
which explains how orbital degeneracy $N$ helps to increase the critical 
$U$ at which the Mott transition takes place and we analyze how frustration 
leads to an asymmetry of the critical $U$ for fillings $n$ and $2N-n$. 
A summary in Sec.\ IV closes the presentation.

\section{Model calculations}

\subsection{Model Hamiltonian}
Solid C$_{60}$ is characterized by a very weak inter-molecular interaction.
Therefore the molecular levels merely broaden into narrow, well separated 
bands.\cite{ldabands} The conduction band originates from the lowest
unoccupied molecular orbital, the 3-fold degenerate $t_{1u}$ orbital.
To get a realistic, yet simple description of the electrons in the 
$t_{1u}$ band, we use a Hubbard-like model that describes the interplay 
between the hopping of the electrons and their mutual Coulomb 
repulsion:\cite{c60mott}
\begin{eqnarray}
 \label{Hamil}
H&=&\sum_{\langle ij\rangle} \sum_{mm'\sigma} t_{im,jm'}\;
              c^\dagger_{im\sigma} c^{\phantom{\dagger}}_{jm'\sigma}\nonumber\\
 &&+U\sum_i\hspace{-1.2ex} \sum_{(m\sigma)<(m'\sigma')}\hspace{-1.6ex}
       n_{im\sigma} n_{im'\sigma'} .
\end{eqnarray}
The sum $\langle ij \rangle$ is over nearest-neighbor sites of an fcc lattice.
The hopping matrix elements $t_{im,jm'}$ between orbital $m$ on molecule $i$
and orbital $m'$ on molecule $j$ are obtained from a tight-binding 
parameterization.\cite{TBparam} The molecules are orientationally 
disordered,\cite{oridisord} and the hopping integrals are chosen such that 
this orientational disorder is included.\cite{hopdisord} The band width for 
the infinite system is $W=0.63\,eV$. The on-site Coulomb interaction is 
$U\approx 1.2\,eV$. 
The model neglects multiplet effects, but we remark that these tend to be
counteracted by the Jahn-Teller effect, which is also not included in the 
model. 

We will investigate the above Hamiltonian
for different integer fillings $n$ of the $t_{1u}$ band. The corresponding 
Hamiltonians describe a hypothetical family of doped Fullerides A$_n$C$_{60}$ 
with space group Fm$\bar{3}$m, i.e.\ an fcc lattice with orientationally 
disordered C$_{60}$ molecules. 
In the calculations we use the on-site Coulomb interaction $U$ as a 
parameter to drive the system across the Mott transition.

\subsection{Quantum Monte Carlo method}
As the criterion for determining the metal-insulator transition we
use the opening of the gap
\begin{equation}\label{Eg}
  E_g = E(N+1) - 2 E(N) + E(N-1) ,
\end{equation}
where $E(N)$ denotes the total energy of a cluster of $N_{mol}$ molecules
with $N$ electrons in the $t_{1u}$ band. Since we are interested in integer
filled systems, $N=n\,N_{mol}$, $n$ an integer.
For calculating the energy gap (\ref{Eg}) we then have to determine ground-state
energies for the Hamiltonian (\ref{Hamil}). This is done using quantum
Monte Carlo.\cite{QMC} Starting from a trial function $|\Psi_T\rangle$ we 
calculate
\begin{equation}\label{proj}
  |\Psi^{(n)}\rangle = [1-\tau(H-w)]^n\;|\Psi_T\rangle ,
\end{equation}
where $w$ is an estimate of the ground-state energy. The $|\Psi^{(n)}\rangle$ 
are guaranteed to converge to the ground state $|\Psi_0\rangle$ of $H$, if 
$\tau$ is sufficiently small and $|\Psi_T\rangle$ is not orthogonal to 
$|\Psi_0\rangle$. Since we are dealing with Fermions, the Monte Carlo 
realization of the projection (\ref{proj}) suffers from the sign-problem. 
To avoid the exponential decay of the signal-to-noise ratio we use
the fixed-node approximation.\cite{QMC} For lattice models this involves
defining an effective Hamiltonian $H_{\rm eff}$ by deleting from $H$ all 
nondiagonal terms that would introduce a sign-flip. Thus, by construction,
$H_{\rm eff}$ is free of the sign-problem. To ensure that the ground-state
energy of $H_{\rm eff}$ is an upper bound of the ground state of the 
original Hamiltonian $H$, for each deleted hopping 
term, an on-site energy is added in the diagonal of $H_{\rm eff}$. Since 
$|\Psi_T\rangle$ is used for importance sampling, $H_{\rm eff}$ depends
on the trial function. Thus, in a fixed-node diffusion Monte Carlo calculation
for a lattice Hamiltonian, we choose a trial function and construct the 
corresponding effective Hamiltonian, for which the ground-state energy 
$E_{\rm FNDMC}$ can then be determined by diffusion Monte Carlo without 
a sign-problem.

For the trial function we make the Gutzwiller Ansatz
\begin{equation}\label{psitrial}
  |\Psi(U_0,g)\rangle = g^D\;|\Phi(U_0)\rangle ,
\end{equation}
where the Gutzwiller factor reflects the Coulomb term 
$U\,D=U\sum n_{i m\sigma} n_{i m'\sigma'}$ in the Hamiltonian (\ref{Hamil}).
$|\Phi(U_0)\rangle$ is a Slater determinant that is constructed by solving
the Hamiltonian in the Hartree-Fock approximation, replacing $U$ by a 
variational parameter $U_0$. Details on the character of such trial functions
and the optimization of Gutzwiller parameters can be found in Ref.\ 
\onlinecite{corrsmpl}.

\noindent
\begin{minipage}{3.375in}
\begin{table}
 \caption[]{\label{lanczos1}
  Total energy (in $eV$) for a cluster of four C$_{60}$ molecules with $6+6$
  electrons (filling 3) for different values of the on-site Coulomb interaction
  $U$. The difference between the fixed-node diffusion Monte Carlo results
  and the exact ground-state energy is shown in the last column. Note that
  $E_{\rm FNDMC}$ is always above the exact energy, as expected for a
  variational method.}  
 \vspace{2ex}
 \begin{tabular}{ddd@{\hspace{1ex}}d} % d@{\hspace{1ex}}d}
  \multicolumn{1}{r}{$U$} &
  \multicolumn{1}{r}{$E_{\rm exact}$} &
  \multicolumn{1}{r}{$E_{\rm FNDMC}$} &
  \multicolumn{1}{r}{$\Delta E$} \\ % &
% \multicolumn{1}{r}{$E_{VMC}$} &
% \multicolumn{1}{r}{$\Delta E$} \\
  \hline
  0.25 &  0.8457 &  0.8458(1)& 0.000 \\ % &  0.8490(2) & 0.003 \\
  0.50 &  4.1999 &  4.2004(1)& 0.001 \\ % &  4.2075(3) & 0.008 \\
  0.75 &  7.4746 &  7.4756(2)& 0.001 \\ % &  7.4873(4) & 0.013 \\
  1.00 & 10.6994 & 10.7004(2)& 0.001 \\ % & 10.7179(5) & 0.019 \\
  1.25 & 13.8860 & 13.8875(3)& 0.002 \\ % & 13.9127(6) & 0.027 \\
  1.50 & 17.0408 & 17.0427(4)& 0.002 \\ % & 17.0728(7) & 0.032 \\
  1.75 & 20.1684 & 20.1711(5)& 0.003 \\ % & 20.2061(4) & 0.038 \\
  2.00 & 23.2732 & 23.2757(10)&0.003 \\ % & 23.3125(6) & 0.039 \\
 \end{tabular}
\end{table}
\end{minipage}

\noindent
\begin{minipage}{3.375in}
\begin{table}
 \caption[]{\label{lanczos2}
  Total energy (in $eV$) for a cluster of four C$_{60}$ molecules with on-site
  Coulomb interaction $U=1\,eV$ for different number of electrons $N_\uparrow
  + N_\downarrow$. The difference between the fixed-node diffusion Monte Carlo
  results and the exact ground-state energy is shown in the last column. 
  $E_{\rm FNDMC}$ is always above the exact energy, as expected for a 
  variational method.}
 \vspace{2ex}
 \begin{tabular}{ccdd@{\hspace{1ex}}d}
  $N_\uparrow$ & $N_\downarrow$ &
  \multicolumn{1}{r}{$E_{\rm exact}$} &
  \multicolumn{1}{r}{$E_{\rm FNDMC}$} &
  \multicolumn{1}{r}{$\Delta E$} \\ 
  \hline
  6 & 5 &  8.4649 &  8.4677(2) & 0.003 \\
  6 & 6 & 10.6994 & 10.7004(2) & 0.001 \\ 
  6 & 7 & 13.3973 & 13.3973(2) & 0.001 \\[1ex]
  8 & 7 & 19.5094 & 19.5109(3) & 0.002 \\
  8 & 8 & 22.9515 & 22.9530(3) & 0.002 \\
  8 & 9 & 26.6590 & 26.6613(3) & 0.002 \\
 \end{tabular}
\end{table}
\end{minipage}

To check the accuracy of the fixed-node approximation, we have 
determined the exact ground-state energies for a (small) cluster of four 
C$_{60}$ molecules using the Lanczos method. For systems with different
on-site Coulomb interaction (Table \ref{lanczos1}) and varying number of
electrons (Table \ref{lanczos2}), we consistently find that the results
of fixed-node diffusion Monte Carlo are only a few $meV$ above the exact
energies.  

\subsection{Quantum Monte Carlo results}
Since the quantum Monte Carlo calculations are for finite clusters of
$N_{mol}$ molecules, we have to extrapolate the calculated energy gaps to
infinite system size. An obvious finite-size effect is the fact that the
one-particle spectrum is discrete, hence there can
be a gap, even for $U=0$. Furthermore, in evaluating (\ref{Eg}), we add and 
subtract one electron to a finite system. Even if we distribute the
extra charge uniformly over all molecules, there will be an electrostatic
contribution of $U/N_{mol}$ to the gap. We therefore introduce
\begin{equation}\label{Egg}
  E_G = E_g - E_g(U=0) - {U\over N_{mol}} .
\end{equation}
These corrections are expected to improve the finite-size extrapolation. 
In practice they turn out to be quite small. For a cluster of 32 C$_{60}$ 
molecules, e.g., $E_g(U=0)$ is typically already less than $10\,meV$.  
In the thermodynamic limit both correction terms vanish, as they should.

\begin{figure}
\vspace*{-0.1in}
\centerline{\resizebox{2.3in}{!}{\includegraphics{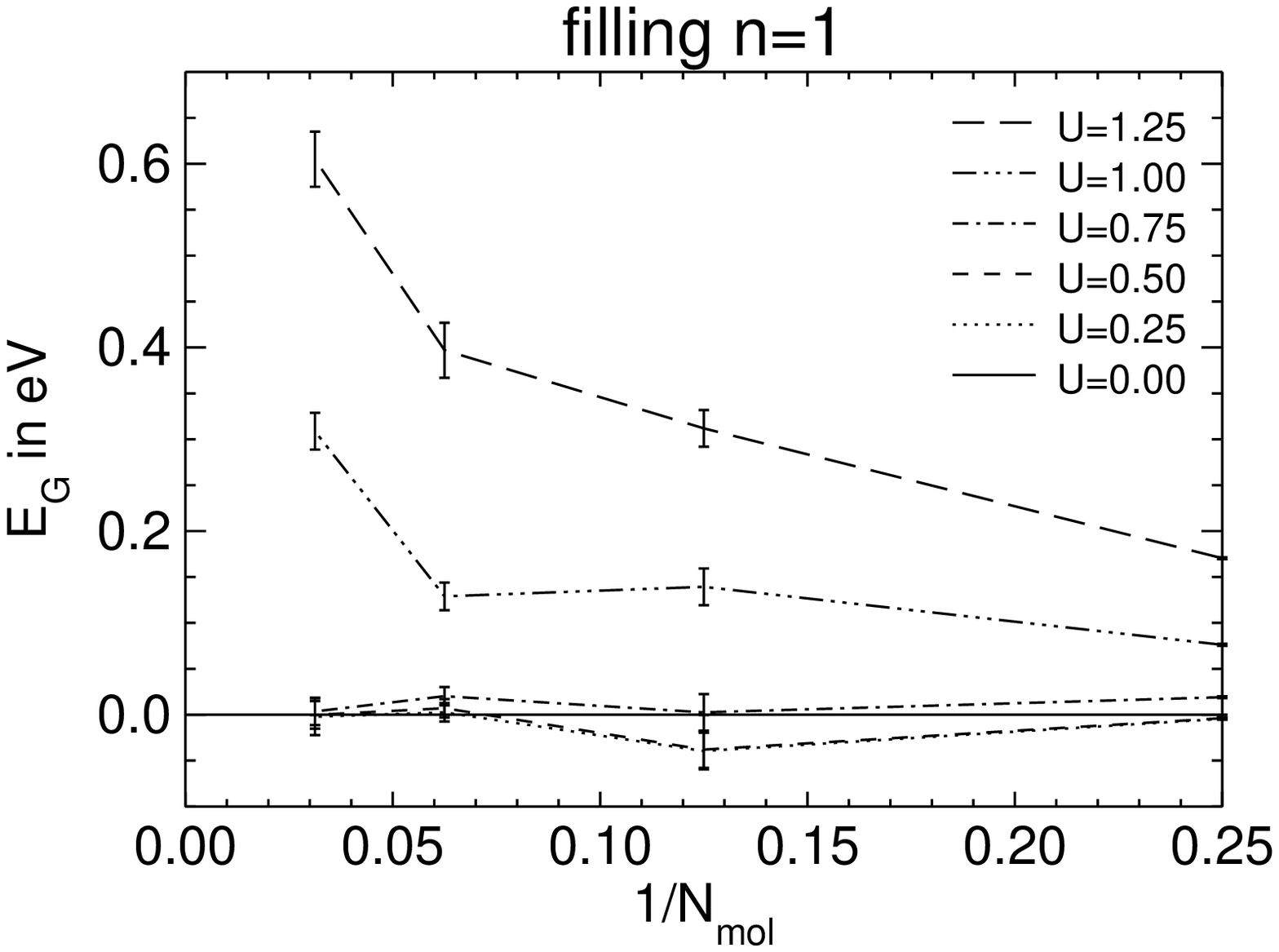}}}
\vspace{-0.05in}
\centerline{\resizebox{2.3in}{!}{\includegraphics{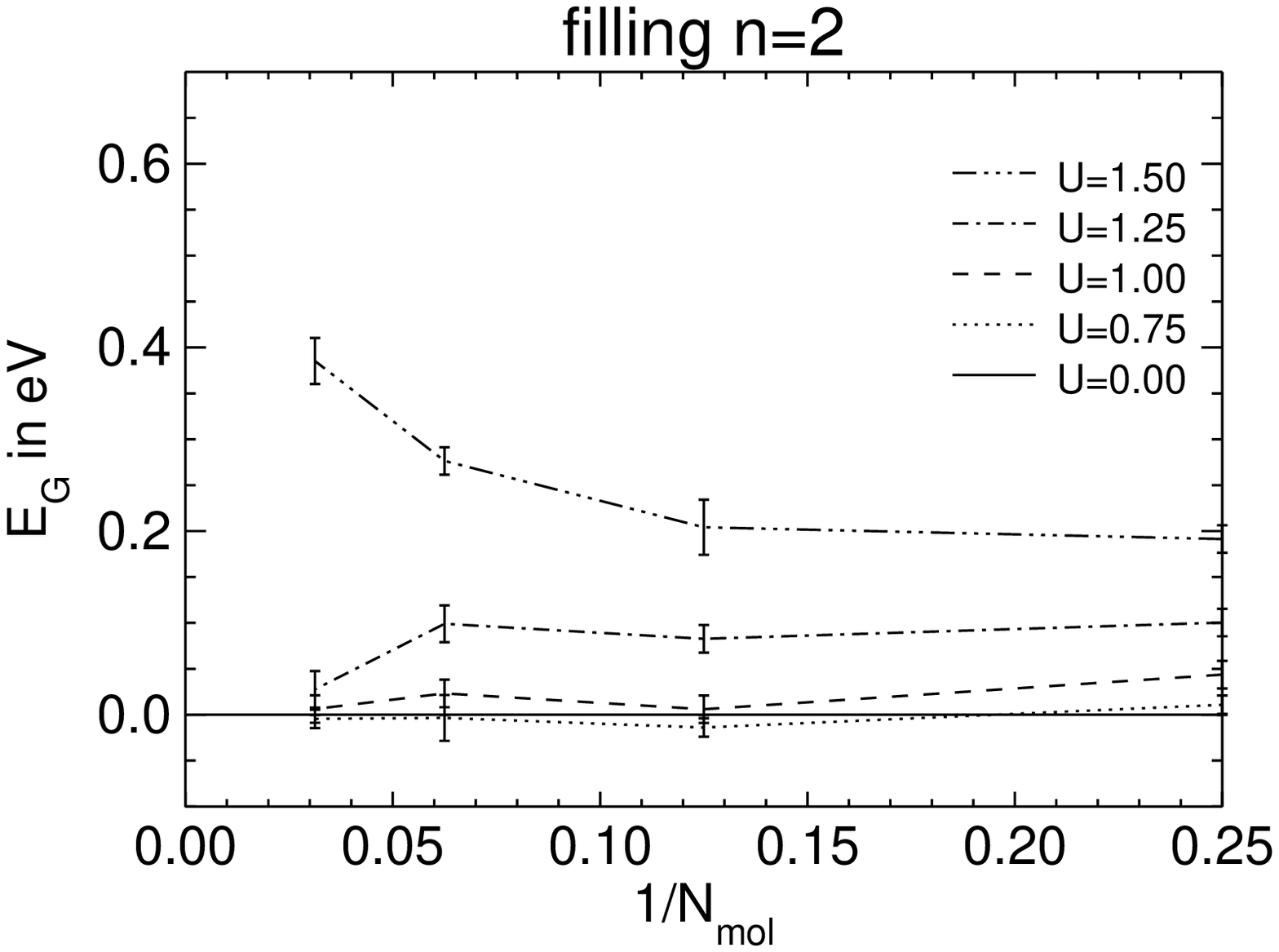}}}
\vspace{-0.05in}
\centerline{\resizebox{2.3in}{!}{\includegraphics{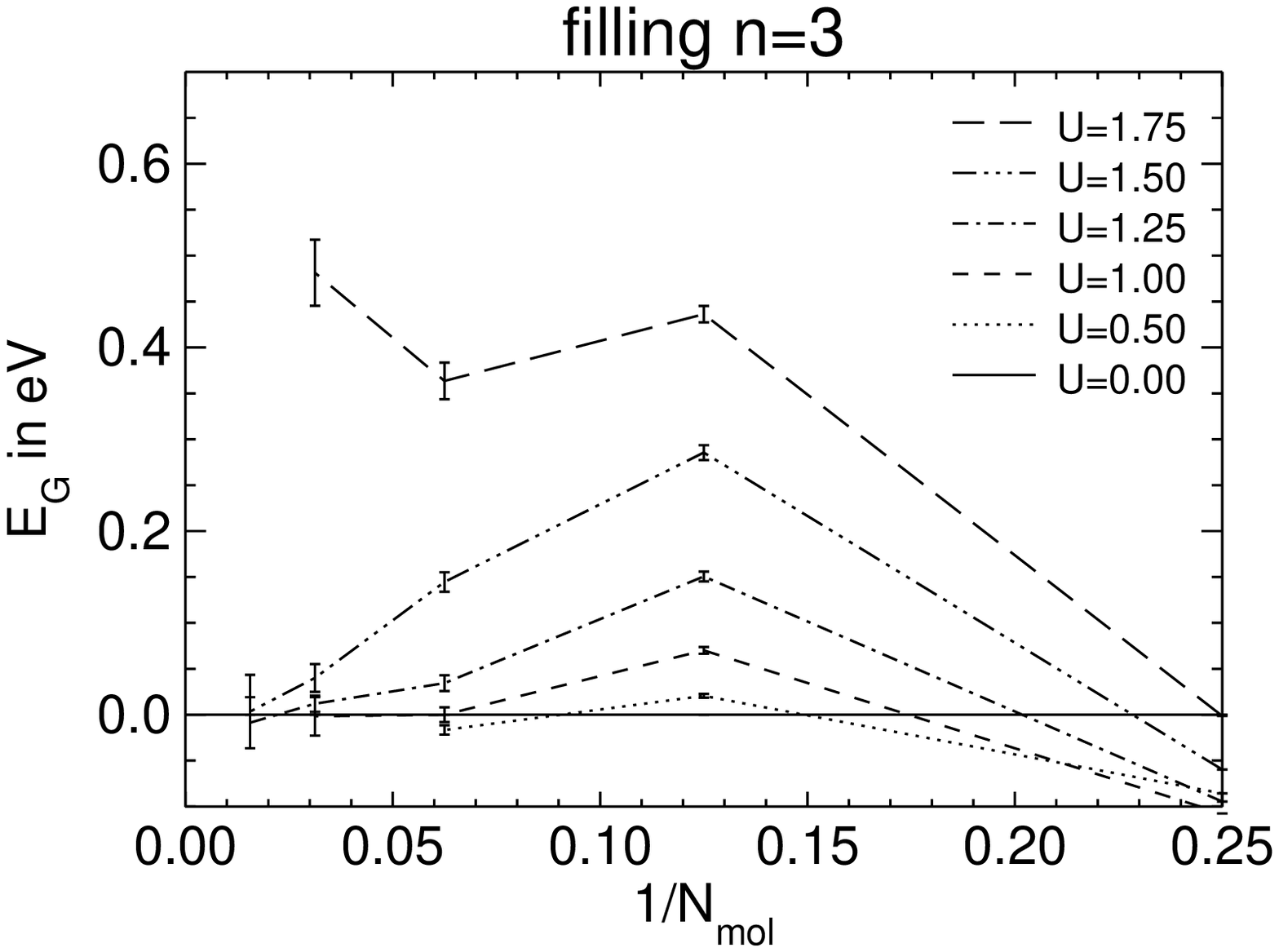}}}
\vspace{-0.05in}
\centerline{\resizebox{2.3in}{!}{\includegraphics{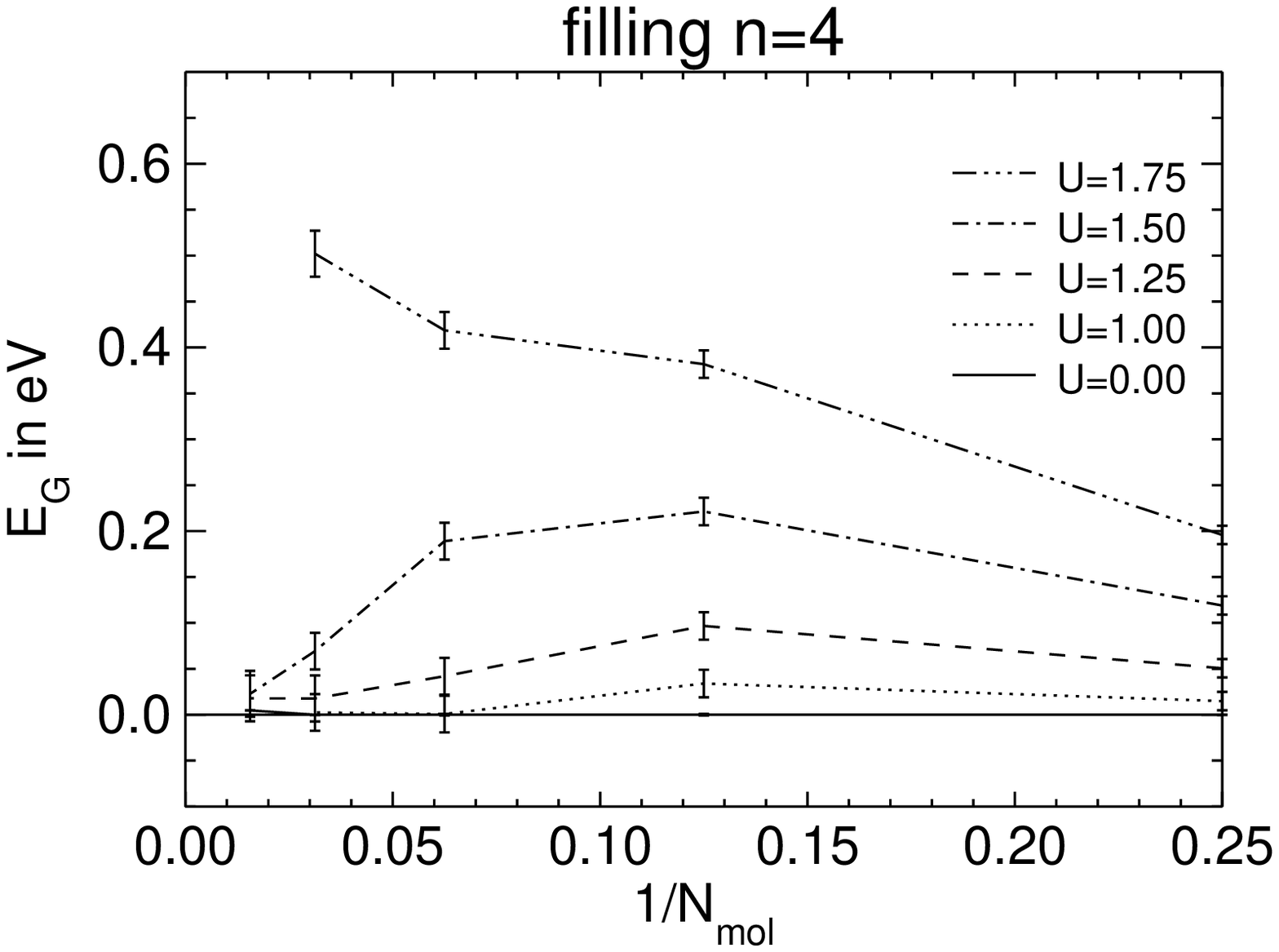}}}
\vspace{-0.05in}
\centerline{\resizebox{2.3in}{!}{\includegraphics{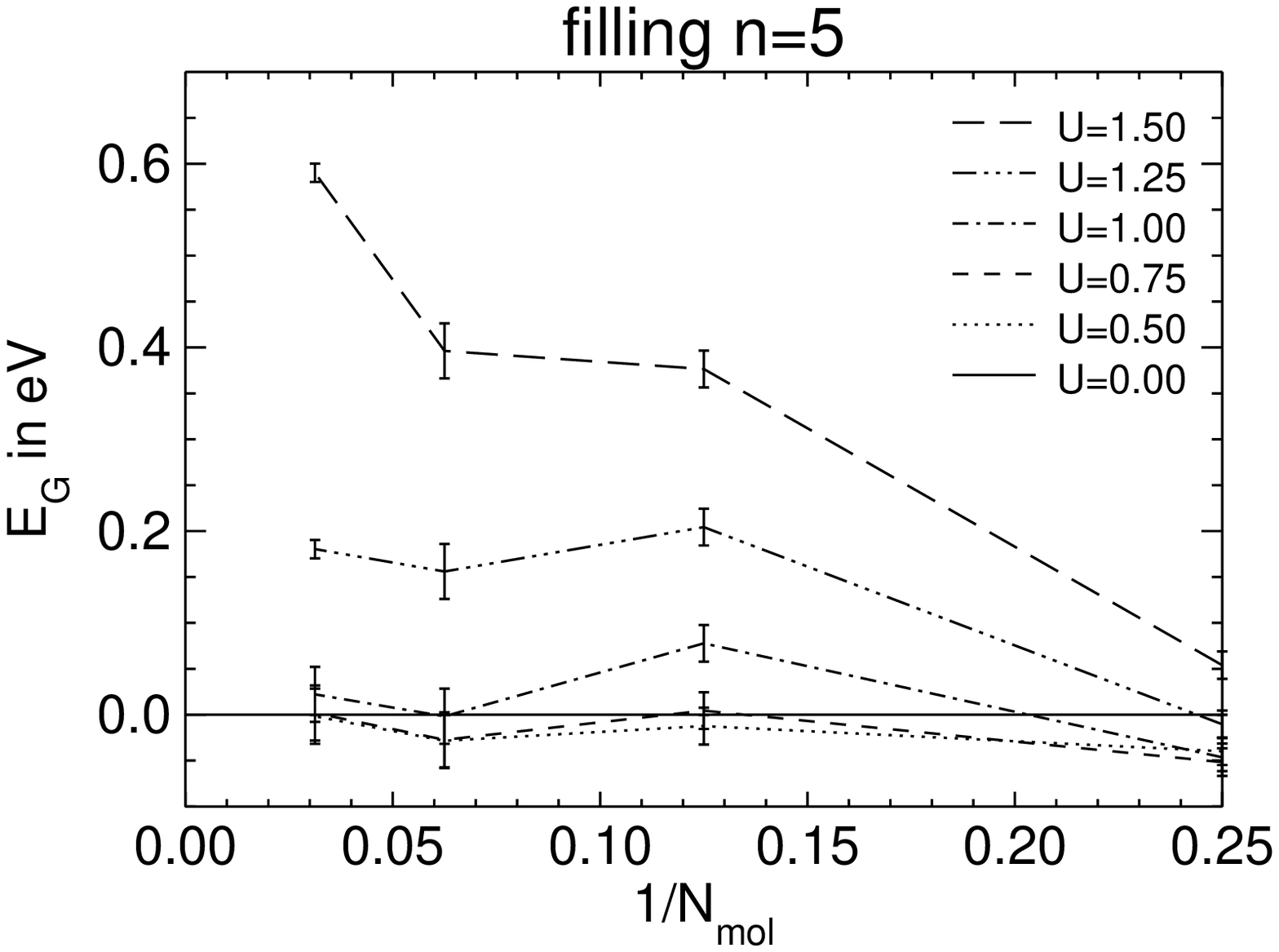}}}
\caption[]{\label{QMCextrap}   
 Finite-size corrected gap (\ref{Egg}) as a function of the inverse number
 of molecules $N_{mol}$ for different values of the Coulomb interaction $U$.
 The error bars give the results of the quantum Monte Carlo calculations for
 systems with $N_{mol}=$ 4, 8, 16, 32, and, where necessary, 64 molecules.
 The lines are merely to guide the eye and identify the value of $U$ in the
 corresponding calculations.}
\end{figure}

\begin{figure}
\centerline{\resizebox{2.8in}{!}{\includegraphics{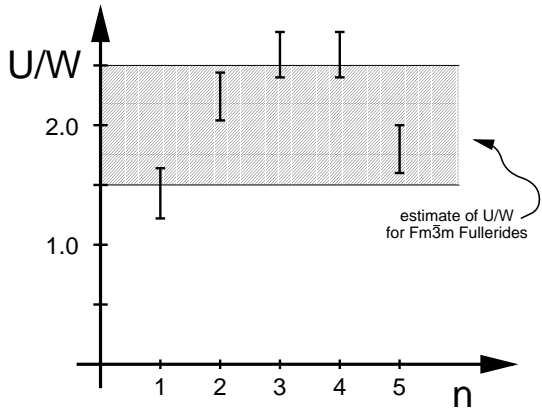}}}
\vspace{2ex}
\caption[]{\label{QMCres}
 Estimate of the critical ratio $U_c/W$ for a multi-band Hubbard model
 describing the doped Fm$\bar{3}$m Fullerides (fcc lattice with orientational
 disorder) at integer fillings $n$.
 The error bars give the estimates for $U_c/W$ from the quantum Monte Carlo
 calculations. The shaded region indicates the $U/W$-range in which the
 real materials are believed to fall.}
\end{figure}

The results of the quantum Monte Carlo calculations are shown in Fig.\ 
\ref{QMCextrap}. Plotting the finite-size corrected gap $E_G$ for different
values of the Coulomb interaction $U$ versus the inverse system size 
$1/N_{mol}$, we read off where the gap starts to open. For the system with
one electron per molecule the gap opens around $U_c\approx0.75\ldots1.00\,eV$.
At filling 2 the transition takes place later, at $U_c\approx1.25\ldots1.50\,
eV$. For both, filling 3 and 4 we find the largest critical $U$: $U_c\approx
1.50\ldots1.75\,eV$. For the system with 5 electrons per molecule the gap
opens around $U_c\approx1.00\ldots1.25\,eV$.
The results are summarized in Fig.\ \ref{QMCres}.
Thus we find that for an otherwise identical Hamiltonian the critical $U$
for the Mott transition depends strongly on the filling. $U_c$ is largest
at half-filling and decreases away from half-filling. The decrease in $U_c$ 
is, however, not symmetric around half-filling. It is more pronounced for 
fillings $<3$ than for fillings $>3$. 

We note that the opening of the gap is accompanied by a change in
the character of that trial function which yields the lowest energy in the 
fixed-node approximation. For small $U$, where the system is still in the
metallic regime, paramagnetic trial functions with small $U_0$ (see the
discussion after eqn.\ (\ref{psitrial})) are best. When the gap starts to open, 
trial functions with larger $U_0$, which have antiferromagnetic character, 
give lower energies. The corresponding Slater determinants 
$|\Phi(U_0)\rangle$ describe a Mott insulator in Hartree-Fock approximation. 

\subsection{Hartree-Fock calculations}
\begin{figure}
\centerline{\resizebox{3.2in}{!}{\rotatebox{270}{\includegraphics{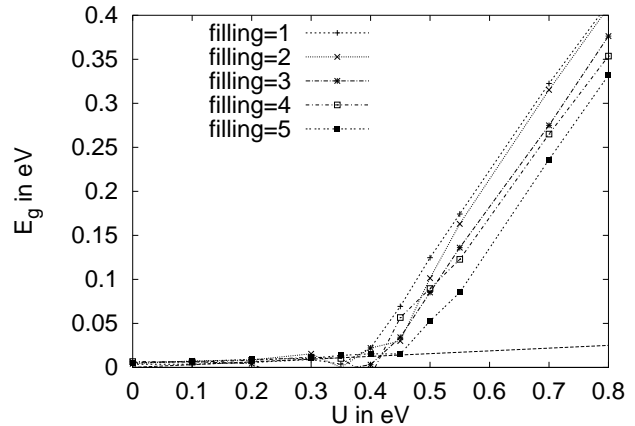}}}}
\vspace{2ex}
\caption[]{\label{HFgap}
           Gap $E_g$ as a function of the Hubbard interaction $U$ for a 
           cluster of 32 molecules in
           Hartree-Fock approximation for integer fillings $n=1\ldots5$. 
           The calculations were done for the same cluster of 32 molecules 
           that was used in the quantum Monte Carlo calculations. The dashed
           line shows the finite-size contribution $U/N_{mol}$ to the gap (cf.\ 
           eqn.\ (\ref{Egg})). The Mott transition occurs around 
           $U\approx 0.4\,eV$ ($U/W\approx 0.65$). $U_c$ depends only weakly 
           on the filling $n$, increasing slightly with increasing $n$.}
\end{figure}

It is instructive to compare the results of the quantum Monte Carlo 
calculations with the predictions of Hartree-Fock theory. Figure \ref{HFgap} 
shows the the gap $E_g$ calculated for the Hamiltonian (\ref{Hamil}) within 
the Hartree-Fock approximation for the different integer fillings. 
Compared with quantum Monte Carlo, the gap opens much too early, around 
$U\approx 0.4\,eV$ ($U/W\approx 0.65$). Furthermore, there is only a very 
weak doping dependence:
$U_c$ somewhat increases with the filling --- in qualitative disagreement
with the quantum Monte Carlo results. This failure is a direct consequence 
of the mean-field approximation. In Hartree-Fock the only way to avoid 
multiple occupancies of the molecules, in order to reduce the Coulomb repulsion,
is to renormalize the on-site energy for the orbitals, thereby localizing the 
electrons in certain orbitals. For the Hamiltonian (\ref{Hamil}) this on-site 
energy is, apart from a trivial offset, given by
$\varepsilon_{im\sigma}=U\langle\sum_{m'\sigma'} n_{im'\sigma'} - 
n_{im\sigma}\rangle$. Lowering the Coulomb energy in this way will, however, 
increase the kinetic energy. For small changes in the on-site energies this 
increase will 
scale like the inverse of the density of states at the Fermi level. This 
suggests that the critical $U$ should be the larger, the smaller the density 
of states at the Fermi level. Inspecting the density of states $N(\varepsilon)$
for the non-interacting Hamiltonian (see e.g.\ Fig.\ 3 of Ref.\ 
\onlinecite{DOS}), we find that this is indeed the case: $N(\varepsilon)$ 
slightly decreases with filling, explaining the corresponding increase
in $U_c$. Hence the weak, but qualitatively wrong, doping dependence in 
Hartree-Fock can be understood as an effect of the small variation in the 
density of states of the non-interacting system. 

\section{Interpretation}\label{degarg}
\subsection{Hopping enhancement}
\begin{figure}
\centerline{\resizebox{1.4in}{!}{\includegraphics{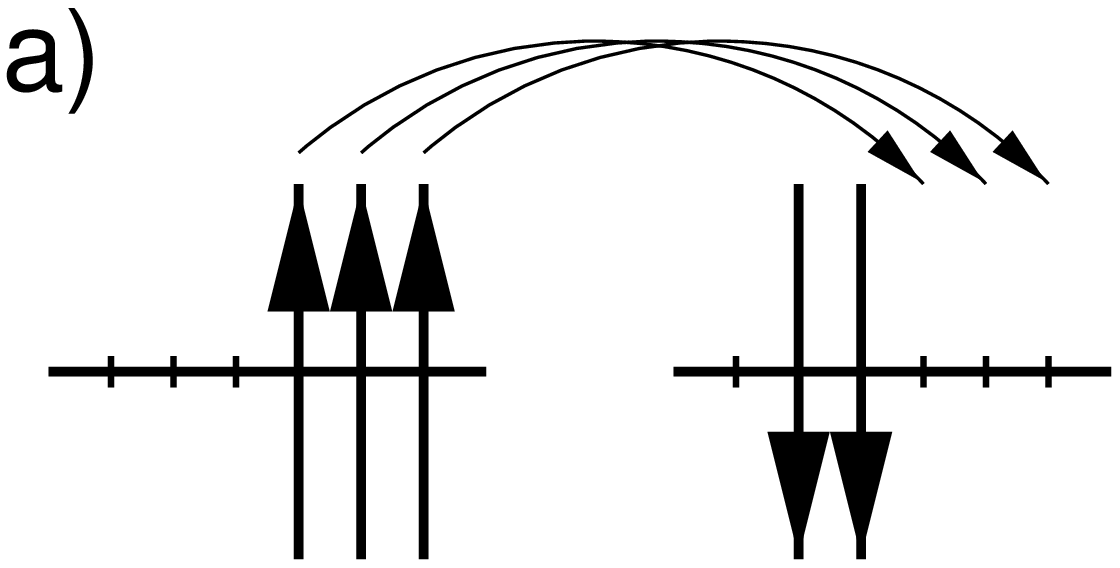}}}
\centerline{\resizebox{1.4in}{!}{\includegraphics{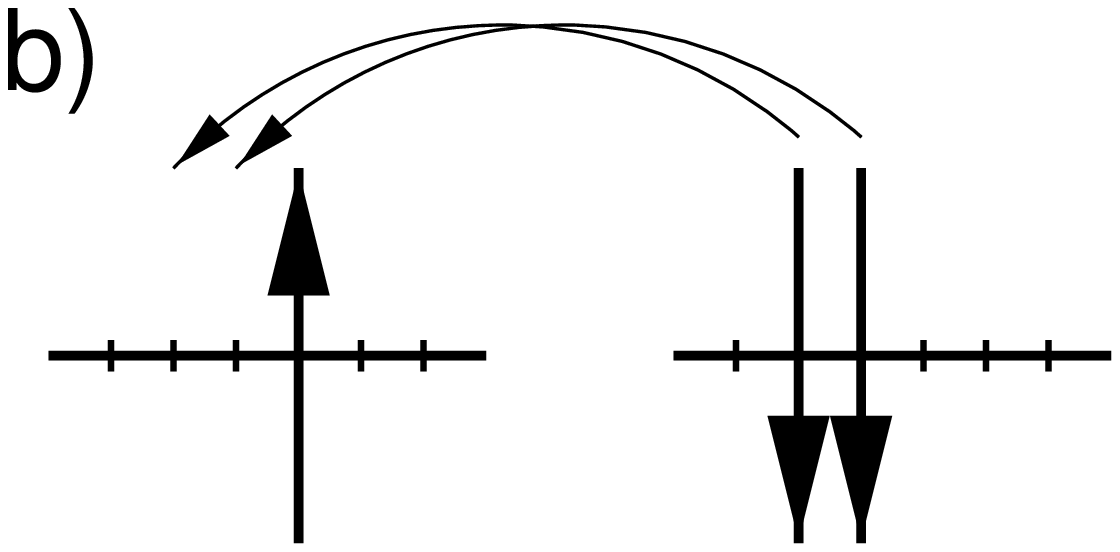}}}
\caption[]{\label{hopping}
 Illustration of how a) an extra electron or b) an extra hole can hop against 
 an integer-filled background (here: degeneracy $N=3$, filling $n=2$). 
 For simplicity\cite{samem} we consider the case where electrons can only hop 
 between orbitals with the same quantum number, 
 i.e.\ $t_{im,jm'}=0$ for $m\ne m'$.}
\end{figure}

To find a simple interpretation for the doping dependence of the Mott 
transition we consider the limit of large Coulomb interaction $U$. In that
limit the Coulomb energy dominates and we can estimate the energies entering
the gap equation (\ref{Eg}) by considering electron configurations in real 
space. According to the Hamiltonian (\ref{Hamil}) the contribution to the
Coulomb energy from a molecule that is occupied by $m$ electrons is 
$U\;m(m-1)/2$. 
Thus the energy of a system with filling $n$ is minimized for configurations 
with exactly $n$ electrons per molecule. The hopping of an electron to a
neighboring molecule would cost the Coulomb energy $U$ and is therefore
strongly suppressed in the large-$U$ limit. The energy for a cluster of
$N_{mol}$ molecules with $N=n\,N_{mol}$ electrons (filling $n$) is then 
given by
\begin{equation}
  E(N)={n(n-1)\over2}\,N_{mol}\,U + {\cal O}(t^2/U) ,
\end{equation}
where $t$ is a typical hopping matrix element. Adding an extra electron
increases the Coulomb energy by $n\,U$, removing an electron reduces
it by $(n-1)\,U$. But there will also be a kinetic contribution to the
energy $E(N\pm1)$, since the extra charge can hop without any additional cost
in Coulomb energy. To estimate the kinetic energy we calculate the 
matrix element for the hopping of the extra charge to a neighboring molecule.
This matrix element will of course depend on the arrangement of the other
$N$ electrons. It is well known that for the non-degenerate Hubbard model
a ferromagnetic arrangement of the spins is energetically favored (Nagaoka's
theorem\cite{Nagaoka}), allowing the extra charge to hop without disturbing 
the background spins. For a degenerate Hubbard model, however, the hopping
matrix element is larger e.g.\ for an antiferromagnetic arrangement of the 
background spins.\cite{samem} This is illustrated in Fig.\ \ref{hopping} a) for
an extra electron in a system with filling 2. Now, instead of only the extra
electron, any one out of the three equivalent electrons can hop to the 
neighboring molecule. Denoting the state with the extra electron on molecule 
$i$ by $|i\rangle$, we find that the second moment of the Hamiltonian 
$\langle i|H^2|i\rangle$ is given by the number of hopping channels $k$ (in the
present case $k=3$) times the number of (equivalent) nearest neighbors $Z$
times the single-electron hopping matrix element $t$ squared. Thus by inserting
the identity in the form $\sum_j |j\rangle\langle j|$, where $|j\rangle$ 
denotes the state where any one of the electrons has hopped form molecule
$i$ to the neighboring molecule $j$, we find
\begin{equation}
  \langle i|H|j\rangle=\sqrt{k}\,t ,
\end{equation}
i.e.\ the hopping matrix element is enhanced by a factor of $\sqrt{k}$ 
over the one-particle hopping matrix element $t$. In a similar way we find
for the system with an extra hole (Fig.\ \ref{hopping} b) a hopping 
enhancement of $\sqrt{k}$ with $k=2$. The hopping enhancements for other
fillings are listed in table \ref{enhancement}, where $k_-$ denotes the
enhancement for a system with an extra hole, and $k_+$ is for a
system with an extra electron. 

\noindent
\begin{minipage}{3.375in}
 \begin{table}
 \caption[]{\label{enhancement}
  Hopping enhancement for different fillings of a 3-fold degenerate band.} 
 \begin{tabular}{lll}
  filling & \multicolumn{2}{c}{enhancement: $\sqrt{k}$}\\
  \hline
   $n=1$  & $k_-=1$  &  $k_+=2$ \\
   $n=2$  & $k_-=2$  &  $k_+=3$ \\
   $n=3$  & $k_-=3$  &  $k_+=3$ \\
   $n=4$  & $k_-=3$  &  $k_+=2$ \\
   $n=5$  & $k_-=2$  &  $k_+=1$
 \end{tabular}
\end{table}
\end{minipage}
%\begin{figure}
%\centerline{\resizebox{2.5in}{!}{\includegraphics{dope.eps}}}
%\vspace{2ex}
%\caption[]{
% Hopping enhancement for different fillings of a 3-fold degenerate band.
% The enhancement factors are $\sqrt{3}\approx 1.75$ for filling 3 
% (half-filling), $(\sqrt{3}+\sqrt{2})/2\approx 1.57$ for fillings 2 and 4,
% and $(\sqrt{2}+1)/2\approx 1.21$ for fillings 1 and 5.}
%\end{figure}
% \begin{tabular}{l@{\hspace{5ex}}c@{$\;\approx\,$}c}
%  filling & \multicolumn{2}{c}{enhancement}\\
%  \hline
%  $n=\;3$ & $\sqrt{3}$                  & 1.73\\[0.5ex]
%  $n=2,4$ & ${\sqrt{3}+\sqrt{2}\over2}$ & 1.57\\[0.5ex]
%  $n=1,5$ & ${\sqrt{2}+   1    \over2}$ & 1.21\\
% \end{tabular}

For a single electron the kinetic energy is of the order of $-W/2$, where $W$
is the one-electron band width. The enhancement factor $\sqrt{k}$ in the 
many-body case then suggests that the kinetic energy for the extra charge is
correspondingly enhanced, implying 
\begin{eqnarray*}
  E(N+1) &\approx& E(N) + \hspace{3ex}n\hspace{3ex}\,U - \sqrt{k_+}\;W/2\\
  E(N-1) &\approx& E(N) - (n-1)\,U - \sqrt{k_-}\;W/2 .
\end{eqnarray*}
Combining these results we find 
\begin{equation}\label{largeUgap}
  E_g \approx U - {\sqrt{k_+}+\sqrt{k_-}\over2}\;W ,
\end{equation}
i.e.\ the hopping enhancement leads to a reduction of the gap described by the 
factor
$c=(\sqrt{k_+}+\sqrt{k_-})/2$. This reduction is largest ($\approx 1.73$) for
n=3, and becomes smaller away from half-filling: $c\approx1.57$ for $n=2,\,4$, 
and $c\approx1.21$ for fillings 1 and 5. Extrapolating (\ref{largeUgap}) to
intermediate $U$ we find that the gap opens for U larger than $U_c=c\,W$.
Therefore the above argument predicts that the critical $U$ for the Mott
transition depends strongly on the filling, with $U_c$ being largest at 
half-filling and decreasing away from half-filling. This is qualitatively
the same behavior as we have found in the Monte Carlo calculations. 
We note, however, that the argument we have presented is not exact. First,
the hopping of an extra charge against an antiferromagnetically ordered
background will leave behind a trace of flipped spins. Therefore the analogy
with the one-electron case for determining the kinetic energy in the large-$U$
limit is not exact. Second, using (\ref{largeUgap}) for determining $U_c$
involves extrapolating the results obtained in the limit of large $U$ to
intermediate values of the Coulomb interaction, where the Mott transition
takes place. Finally, considering only one nearest neighbor in the hopping
argument (cf.\ Fig.\ \ref{hopping}) implicitly assumes that we are dealing
with a bipartite lattice, where all nearest neighbors are equivalent.

\subsection{Origin of the asymmetry}

To analyze the asymmetry in the gaps around half-filling we use the following
exact relation for the kinetic energy in the limit of infinite $U$, which
follows from an electron-hole transformation
\begin{equation}\label{symm}
  T_{min\atop max}(nN_{mol}\pm1)=-T_{max\atop min}((2N-n)N_{mol}\mp1) .
\end{equation}
(Note how this symmetry is reflected in the hopping enhancements shown
in Table \ref{enhancement}.)
Since the gap for filling $n$ is given by
\begin{displaymath}
  E_g(n)=U+T_{min}(nN_{mol}-1)+T_{min}(nN_{mol}+1) ,
\end{displaymath}
the asymmetry $\Delta=E_g(n)-E_g(2N-n)$ in the gaps can be written 
entirely in terms of energies for systems with an extra electron:
\begin{displaymath}
  \Delta = \begin{array}{ccc}
             -T_{max}((2N-n)N_{mol}+1) &+& T_{min}(nN_{mol}+1)\\
             -T_{min}((2N-n)N_{mol}+1) &+& T_{max}(nN_{mol}+1) .
           \end{array}
\end{displaymath}
For a bipartite system the spectrum for a given filling will be symmetric,
in particular $T_{min}+T_{max}=0$, and thus there will be no asymmetry
in the gaps: $\Delta=0$. 
Frustration breaks this symmetry. To study the effect of
frustration we perform a Lanczos calculation in the large-$U$ limit, 
starting from a configuration $|v_0\rangle$ of the type shown in 
Fig.\ \ref{hopping}. The leading effect of frustration is given by the 
third moment, which already enters after the first Lanczos step. 
Diagonalizing the Lanczos matrix and expressing everything in terms of 
the moments of the Hamiltonian, the extreme eigenvalues are given 
by\cite{bwidth}
\begin{equation}\label{Lanczos}
  \varepsilon_{max\atop min}={A_3\pm\sqrt{4A_2^3+A_3^2}\over2A_2} ,
\end{equation}
where $A_k=\langle v_0|H^k|v_0\rangle$ denotes the $k^{th}$ moment of $H$,
and $A_1=\langle v_0|H|v_0\rangle=0$ for a state like in Fig.\ \ref{hopping}. 
From this expression it is clear that the ``band width'' $\varepsilon_{max}
-\varepsilon_{min}$ is essentially given by the second moment, and that
an enhancement of $A_2$ by a factor of $k$ leads to an increase in the band
width by a factor of $\sqrt{k}$, as already described above. The main effect 
of the third moment (i.e.\ of frustration) is to shift the extremal 
eigenvalues, where the shift is determined by the third moment. 

To get a contribution to the third moment the initial state $|v_0\rangle$ 
must be recovered after three hops. 
%%%%This is only possible if one electron hops around a triangle.
This is only possible if the extra
charge hops around a triangle, without changing spins along its path.
For a state with an extra electron this means that {\em one and the same}
electron has to perform the triangular hop.
Therefore, even in the many-body case, for each such electron we get
the same contribution to the third moment as in the single electron case. 
It therefore makes sense to write the third moment $A_3(n)$ for a system with 
$nN_{mol}+1$ electrons in terms of the third moment $A^s_3$ of the single 
electron problem: $A_3(n)=\kappa_+(n)\,A^s_3$, where $\kappa_+(n)$ describes
the many-body effects, just like we introduced $k_+(n)$ to describe the 
many-body enhancement of the second moment. Using these definitions we find
that the size of the asymmetry $\Delta$ in the gaps can be estimated by the 
doping dependence of the (positive) enhancement factors $\kappa_+(n)$ of the 
third moment, while the overall sign is determined by the single-electron 
moments:
\begin{equation}\label{asym}
  \Delta \approx \left( {\kappa_+(n)   \over k_+(n)   }
                       -{\kappa_+(2N-n)\over k_+(2N-n)}\right)\;
                 {A^s_3\over A^s_2} .
\end{equation}

\begin{figure}
 \centerline{\resizebox{2.8in}{!}{\includegraphics{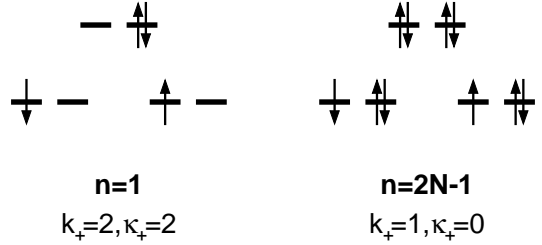}}}
 \vspace{2ex}
 \caption[]{\label{tri2}
            Typical states for a triangle with orbital degeneracy $N=2$
            and hopping only between like orbitals. For filling $n=1$ there is
            no reduction of frustration ($\kappa_+=k_+$). For filling larger
            than $N$ the ``background electrons'' block the triangular
            moves, suppressing frustration.}
\end{figure}

To understand the doping dependence of $\kappa_+/k_+$ we proceed in two steps.
First we observe that the upper limit for the number of different electrons 
that can perform a triangular hop is given by the number $k_+$ of electrons 
that can hop to a nearest neighbor. Hence, if frustration is not suppressed,
$\kappa_+/k_+ = 1$. For filling $n=1$, $N\ge2$ this upper limit can always be 
achieved without compromising large next-neighbor hopping by arranging the 
electrons in such a way as to avoid each other. This is shown in 
Fig.\ \ref{tri2}.
For the corresponding filling $2N-1$ the electrons can no longer be completely 
separated in that way. Thus the channel for triangular hops will be blocked 
by the the Pauli principle, reducing $\kappa_+/k_+$. 
In that way for the larger fillings frustration is reduced. 

This reduction of frustration can already be seen in the simple model of a 
triangle with orbital degeneracy $N=2$ (cf.\ Fig.\ \ref{tri2}). Choosing 
matrix elements $t=1$ for hopping between like orbitals we find for filling 
$n=1$ a strong asymmetry $T_{min}(3n+1)=-2$ and $T_{max}(3n+1)=+4$, while at
filling $n'=2N-n=3$ there is no asymmetry in the extremal eigenvalues:
$T_{max\atop min}(3n'+1)=\pm2$. We note that flipping one spin in the
configuration for filling $n=3$ would allow for a triangular hop. In a
Lanczos calculation this spin polarized configuration gives, however,
only extremal eigenvalues $T_{min}=-2$ and $T_{max}=+1$. The states described
here for a triangle can be easily adapted to the situation in an fcc lattice,
where the third moment involves hopping to the nearest neighbor sites, which
form connected triangles.
%\noindent
%\begin{minipage}{3.375in}
%\begin{table}
% \begin{tabular}{ccc}
%  filling & $E_{min}$ & $E_{max}$ \\
%  \hline
%    n=1   &  -2        &   4      \\
%    n=2   &  -3        &   3      \\
%    n=3   &  -2        &   2      \\
% \end{tabular}
% \caption[]{Extremal eigenvalues of states with an extra electron for 
%            a triangle with orbital degeneracy $N=2$ and matrix elements
%            $t=1$ for hopping between like orbitals.}
%\end{table}
%\end{minipage}

From the non-interacting density of states for our model of the doped 
Fullerenes (cf.\ e.g.\ Fig.~3 of Ref.~\onlinecite{DOS}) we see that both 
$\varepsilon_{min}$ and 
$\varepsilon_{max}$ are shifted upwards, compared to the center of the band,
hence, looking at (\ref{Lanczos}) we find that for a single electron the 
third moment is positive: $A^s_3>0$. Together with the reduction of the 
frustration for larger filling, we therefore expect from (\ref{asym}) that
for the alkali doped Fullerenes $E_g(n)>E_g(2N-n)$; i.e.\ $U_c(n)<U_c(2N-n)$,
as is found in the Monte Carlo calculations.

\section{Summary}

Using quantum Monte Carlo, we have analyzed a model of alkali-doped Fullerenes
and found that the Mott transition strongly depends on the (integer) filling 
$n$.  $U_c$ is largest for $n=3$ and decreases away from half-filling. This 
result is qualitatively different from both, the results of density functional
calculations in the local density approximation, and the results of
Hartree-Fock calculations. The doping dependence of the Mott transition can
be understood in terms of a simple hopping argument. The key observation is
that, due to the orbital degeneracy, there are more hopping channels in the 
many-body than in the single-body case, thus leading to the degeneracy 
enhancement $\sqrt{k}$ discussed above. In addition, due to frustration,
the gaps are not symmetric around half-filling.

The Gutzwiller approximation for a paramagnetic state also predicts a 
degeneracy enhancement.\cite{GA} For a half-filled system, the predicted 
enhancement is, however, linear in the degeneracy $(N+1)$ instead of 
$\sqrt{N}$ as suggested by the hopping argument of Sec.\ \ref{degarg} and 
as also found in infinite dimensions.\cite{jong} 
%%%%It is, furthermore, not clear
%%%%what the physical meaning of the results of the Gutzwiller approximation is,
%%%%since it is an approximation to the paramagnetic Gutzwiller wavefunction,
%%%%which most likely does not exhibit a Mott transition at all.
The results of the Gutzwiller approximation are reproduced by a slave-boson 
calculation in the saddle-point approximation.\cite{SB} 
In dynamical mean-field theory a degeneracy enhancement and a reduction of
$U_c$ away from half-filling, similar to our result, is found.\cite{jong,infd}
%%%%Exactly identifying the Mott transition in such calculations seems, however,
%%%%to be difficult.\cite{doubts}

\section*{Acknowledgments}
This work has been supported by the Alexander-von-Humboldt-Stiftung under the
Feodor-Lynen-Program and the Max-Planck-Forschungspreis, and by the Department
of Energy, grant DEFG 02-96ER45439.

\end{multicols}

\begin{thebibliography}{*}
\bibitem{samem}
  Here we consider the case where there is only hopping between the same
  orbitals on different sites. The argument can also be made for the case
  where hopping is also allowed between different orbitals. The resulting
  hopping enhancement factors $\sqrt{k}$ relative to the single-electron case
  are the same.
\bibitem{bwidth}
  We note that eqn.\ (\ref{Lanczos}) seems to imply that frustration also
  slightly increases the band width. It is, however, unlikely that this small
  effect will survive in the full calculation.
\end{thebibliography}
\end{document}